\documentclass[letterpaper,twocolumn,10pt]{article}
\usepackage{usenix2021_SOUPS}

\usepackage{tikz}
\usepackage{amsmath}
\usepackage{amssymb}

\usepackage[utf8]{inputenc}
\usepackage{graphicx}
\usepackage{tikz}
\usetikzlibrary{calc,positioning,arrows}
\usepackage{verbatim}
\usepackage{fancyvrb}
\usepackage{url}
\usepackage{xcolor}
\usepackage{todonotes}
\usepackage{float}
\usepackage{framed}
\usepackage{soul}
\usepackage{textcomp}

\usepackage[utf8]{inputenc}

\usepackage{todonotes}
\presetkeys{todonotes}{fancyline, color=blue!30,size=\tiny}{}

\newcommand{\PVV}{\textit{PVV}} 

\begin{document}
%-------------------------------------------------------------------------------

%don't want date printed
\date{}

\title{Phrase-Verified Voting:\\
Verifiable Low-Tech Remote Boardroom Voting\\[8pt]
{\it (How We Voted on Tenure \& Promotion Cases during the Pandemic)}}

% if you leave this blank it will default to a possibly ugly attempt 
% to make the contents of the \author command below into a string
\def\plainauthor{}

%for single author (just remove % characters)
\author{
{\rm Enka Blanchard}\\
Digitrust, Loria, \\ Universit\'{e} de Lorraine 
\and
{\rm Ryan Robucci}\\
Cyber Defense Lab\\University of Maryland, Baltimore County (UMBC)
\and
{\rm Ted Selker}\\
Cyber Defense Lab (UMBC)\\Selker Design Research
\and
{\rm  Alan T. Sherman}\\
Cyber Defense Lab\\University of Maryland, Baltimore County (UMBC)
}
%%%%%%%%%%%%%%%%%%%%%%%%%%%%%%%%%%%%%%%%%%%%%%%%%%%%%%%%%%%%%%%%%%%%%%%%%%%
%\author{{\rm Anonymous authors}\\
%Redacted Institutions
%} % this redefines authors for the anonymous publication (we'll remove it when/if the paper gets accepted)
\maketitle
\thecopyright
\begin{abstract}

We present  {\it Phrase-Verified Voting}, a voter-verifiable remote voting system assembled from commercial off-the-shelf software for small private elections.  
The system is transparent and enables each voter to verify that the tally includes their ballot selection without requiring any understanding of cryptography.
This paper describes the system and its use in fall 2020, to vote remotely in promotion committees in a university.

Each voter fills out a form in the cloud with their vote $V$ (YES, NO, ABSTAIN) and a passphrase $P$---two words entered by the voter.
The system generates a verification prompt of the $(P,V)$ pairs and a tally of the votes, organized to help visualize how the votes add up.
After the polls close, each voter verifies that this table lists their $(P,V)$ pair and that the tally is computed correctly.

The system is especially appropriate for any small group making sensitive decisions.
Because the system would not prevent a coercer from demanding that their victim use a specified passphrase, it is not designed for applications where such malfeasance would be likely or go undetected.

Results from  43 voters show that the system was well-accepted, performed effectively for its intended purpose, and introduced users to the concept of voter-verified elections.
Compared to the commonly-used alternatives of paper ballots or voting by email, voters found the system easier to use, and that it provided greater privacy and outcome integrity.

\end{abstract}

\begin{comment}
Keywords.
Phrase-Verified Voting, 
boardroom voting, remote voting, low-tech voting
\end{comment}

%%%%%%%%%%%%%%%%%%%%%%%%%%%%%%%%%%%%%%%%%%%%%%%%%%%%%%%%%%%%%%%%%%%%%%%%%%%
\section{Introduction}
\label{sec:intro}
The COVID-19 pandemic prompted many universities to carry out their operations remotely.
In normal times, promotion committees follow boardroom voting protocols: they meet, deliberate, and vote in person. This work resulted from a design challenge to help a university that cares about such meetings having a secret ballot when everyone is voting from virtual platforms during the pandemic lockdown.
We describe \textit{Phrase-Verified Voting (PVV)},  designed to have more outcome integrity and ballot privacy than the existing remote voting approach of sending votes by email. 

In this context, the chair of the Promotion Committee in the Computer Science and Electrical Engineering  Department of the University of Maryland, Baltimore County asked for a recommendation on how voting should take place remotely in fall 2020. The system needed to require few steps, be learned and used in a few minutes, and be compatible with the institutional constraints of the university.
We designed, implemented, and fielded \PVV{} as a remote ``boardroom voting system,'' using a common cloud-based form.

In some public referenda---such as most votes in parliamentary houses---individual votes are public, and some universities use public voting as part of the deliberation around promotion cases. Many other university promotion systems and decisions made by  boards of directors require ballot privacy.  A traditional ``boardroom election'' is a referendum voted on by a small number of people (university faculty for instance) who can all fit into one room, where they can see and hear each other~\cite{blanchard2020boardroom}. There are many examples of boardroom elections with sensitive outcomes, from  business directions to  home owners associations. 

Figure~\ref{fig:verify} illustrates the verification prompt---the keystone of \PVV{}.
Voters verify this prompt, which lists each vote together with a passphrase that enables the voter to verify that their vote is included in the tally.

As a potential replacement for private boardroom voting approaches, \PVV{} provides several improvements over email voting (and possibly improvements over in-person too).  
It includes a structure to record the list of who could and did vote, to collect votes automatically, and to assure voters that their vote is included in the tally. 
\PVV{} is designed with commercial off-the-shelf systems for cases where elections are run remotely, where organizers also want to ensure outcome integrity, and that the tally accurately reflects voter intentions. Unlike many online voting systems such as Helios, \PVV{} does not use mathematical cryptography and does not require advanced knowledge of modern security tools.
The system can be used as a teaching tool to introduce the value of verifiable voting.

Our contributions include:
\begin{itemize}
\item a remote voting system for giving remote boardroom voting verification that does not depend on software, retains voter privacy, and prevents the election authorities from tampering with election outcomes.

\item a demonstration that stakeholders considered it advantageous, with two voter groups using it a total of 14 times. 
\end{itemize}

\begin{figure}
    \centering

\begin{tabular}{ll}
{\bf Item} & {\bf Passphrase} \\[8pt]
 YES: & \\ \hline
1. & assume jockey \\
2. & disagree imperial \\
3. & friendly, root \\[8pt]
NO: & \\ \hline
1. & frank 99 \\
2. & presidential shock \\[8pt]
ABSTAIN: & \\ \hline
1. & k b \\[8pt]
Tally & \\ \hline
YES: & 3 \\
NO: & 2 \\
ABSTAIN: & 1 \\
\end{tabular}

    \caption{Example of a formatted verification prompt, which 
    contains organized and sorted data to facilitate
    a voter to check the tally and its inclusion of their vote. Two voters did not strictly respect the ``two words'' instruction.}
    \label{fig:verify}
\end{figure} 

\section{Background and Previous Work} 
\label{sec:backandprevious}

Many important decisions take place in boardrooms, in-person and remotely, including decisions about polices, strategies and leadership and constituency, with a variety of available voting options.
In our case, university promotion committees had been voting in person with paper ballots, but had switched to online video meetings.
By contrast, promotion committees in some other schools typically vote by 
voice, choosing to expose voter choices with no privacy.
In fall 2020, most promotion committees at the university in question voted by email, with members sending their ballot selections to a trusted third party.

%\footnote{Private correspondence with Ronald Rivest, 2020.} 

Desirable properties of voting systems include outcome integrity, ballot privacy, the ability to learn and use it in a few minutes, reliable operations, and an architecture simple enough to deploy and analyze reliably.   
Multiple solutions have been proposed for cases similar to the one at hand; Section~\ref{sec:alternatives}
describes the decision process between key alternatives.

Many solutions are based on advanced cryptography. 
Hao and Ryan~\cite{hao2016real} survey modern electronic voting systems, including end-to-end (E2E) verifiable systems such as Scantegrity~II~\cite{carback2010scantegrity}.
Chaum et al. of the VoteXX Project~\cite{votexx} designed a remote voting system that is coercion-resistant.  Several systems are also specifically tailored for verifiable cryptographic remote boardroom voting---e.g., Javani and Sherman~\cite{javani2020bvot}. 
We also considered
Helios, a free open-source electronic voting system that offers voter verifiability with quite strong security properties, as well as its derivatives~\cite{adida2008helios,adida2009electing,Cortier2019Beleniossimpleprivate}.

% Among those systems, 
% , was also considered - please avoid awkward passive voice

On the other hand, low-tech solutions, not depending on computer programs---or at least complex cryptography---can offer advantages: they do not require complex analysis and have fewer steps to understand and implement. 
The simplest such system is voice voting, which offers high outcome integrity but no ballot privacy. 
The initial email voting used by the university in question offered no voter verifiability, mixed votes with other email communications, and required complete trust in a third party for ballot privacy and outcome integrity.
 
Blanchard, Selker, and Sherman~\cite{blanchard2020boardroom} proposed systems for in-person low-tech verifiable boardroom voting, and some video conference services offer non-verified remote voting (e.g., Zoom or BigBlueButton). 
To our knowledge, this paper proposes the first low-tech verifiable procedure for remote boardroom voting. 

%%%%%%%%%%%%%%%%%%%%%%%%%%%%%%%%%%%%%%%%%%%%%%%%%%%%%%%%%%%%%%%%%%%%%%%%%%%
\section{Problem Specification and Adversarial Model}
\label{sec:probandmodel} 

We {sought} a method to demonstrate that a university department's promotion committee can increase confidence in remotely held votes. 
The challenge was how accuracy, integrity, and privacy could all be improved, in comparison to emailing votes, for a remote boardroom voting experience in a short time without large resources.

The method {should} not add significant effort, time, or expense to the process.
The goals {are} to:
\begin{enumerate}
    \item ensure that the outcome is accurate;
    \item enable voters to verify the correctness of the outcome and that their votes are correctly recorded and tabulated;
    \item provide ballot secrecy;
    \item produce a list of people who voted and a list of eligible voters who did not vote;
    \item require few steps, to ensure that it is easy to learn to use and administer;
    \item be acceptable to the organization holding the vote~\cite{umbchandbook}.
\end{enumerate}

The approximately 35 eligible voters who {used it were} mainly the tenured faculty in a department; each holds a PhD in computer science or a related field. 
Because the stakeholders {are} known and accountable, integrity {is} the most important concern; the method {will have to improve} assurance of a valid outcome even if it {does not have to provide} an extremely high level of security. 
It {is very important to collect} the votes with few steps and without complicating the promotion process. 

%? can we be more descriptive around why or how security  isn't so important 
%it can be 14 or more years of work PhD and tenure quest on the line.

There {are} a multitude of referenda related to personnel in the course of a typical academic year related to promotion and tenure.  For each  promotion candidate, there {are} separate referenda for teaching, service, research, and overall. The committee votes on these referenda in a series of meetings across months, and a few different candidates might be considered in any one meeting. Each meeting lasts approximately one hour.
Each ballot choice for a referendum is one of YES, NO, or ABSTAIN. Voting ABSTAIN is different from not voting. Voters may vote during a committee meeting, or, with permission they may vote by absentee ballot before the meeting.
All eligible voters not on sabbatical or other approved leave are required to vote.

The adversary {could be} anyone, including voters, staff, candidates, and administrators. The goals of the adversary are to modify the outcome or to learn how certain individuals voted. We assume that the adversary is covert in the sense that they do not wish to be caught and do not want people to know that there was an attack. 

We do not consider undue influence attacks (including voter coercion and vote buying) or discreditation attacks that aim to cast doubt on the validity of the outcome.

If needed, the system may use a trusted party (e.g., ideally, an agent who does not know or have interest in the outcome), for tasks which require privacy. 

\section{Alternatives} 
\label{sec:alternatives}

We considered three competing alternative voting systems:
\begin{enumerate}

\item (Baseline) With essentially no assurance of outcome integrity, each voter emails their vote to a trusted third party who accumulates the emails to tabulate their contents.\footnote{A higher-integrity variation is to email the votes to two separate trusted parties, each of whom tabulates them independently.}

\item (Helios) Use the Helios verifiable online voting system, which is available through a web interface~\cite{helios,adida2008helios,adida2009electing}.
Helios is a free, open-source electronic voting system that offers voter verifiability with quite strong security properties~\cite{helios-attacks-and-defense,smyth2018ballot,cardillo2018threat,chang2016cloudier}.

\item (Public Vote) With high outcome integrity and no privacy, each voter fills in a form that generates a public spreadsheet that lists, for each voter, their name and vote. 

\end{enumerate}

The department wanted to achieve higher outcome integrity than offered by Baseline, and required ballot privacy not offered by Public Vote. 

Although Helios provides integrity, security, and accuracy, and is fairly easy to use, it presented four concerns.
First, it would have required special authorization from  the university administration, given that
the servers do not run on the university network. 
Second, considerable time and effort might be required to set it up and run it (especially on the university's servers).
Third, it is more complicated, partially because understanding the inner workings requires some technical expertise---which can create problems for voters, even if voting itself can be simple, thanks to a usable interface.
Fourth, it would have required introducing people to a new system, since faculty have experience using and administering the chosen forms system but none had experience using and administering Helios.  
Nevertheless, Helios should be used for those who seek greater outcome integrity and are willing to administer a more complicated system.

%, especially for elections including over 100 voters.
% boardroom voting usually is <= about 40
% maybe somewhere else we should discuss limit on voters for PVV

%%%%%%%%%%%%%%%%%%%%%%%%%%%%%%%%%%%%%%%%%%%%%%%%%%%%%%%%%%%%%%%%%%%%%%%%%%%

\begin{figure}
    \centering
    \includegraphics{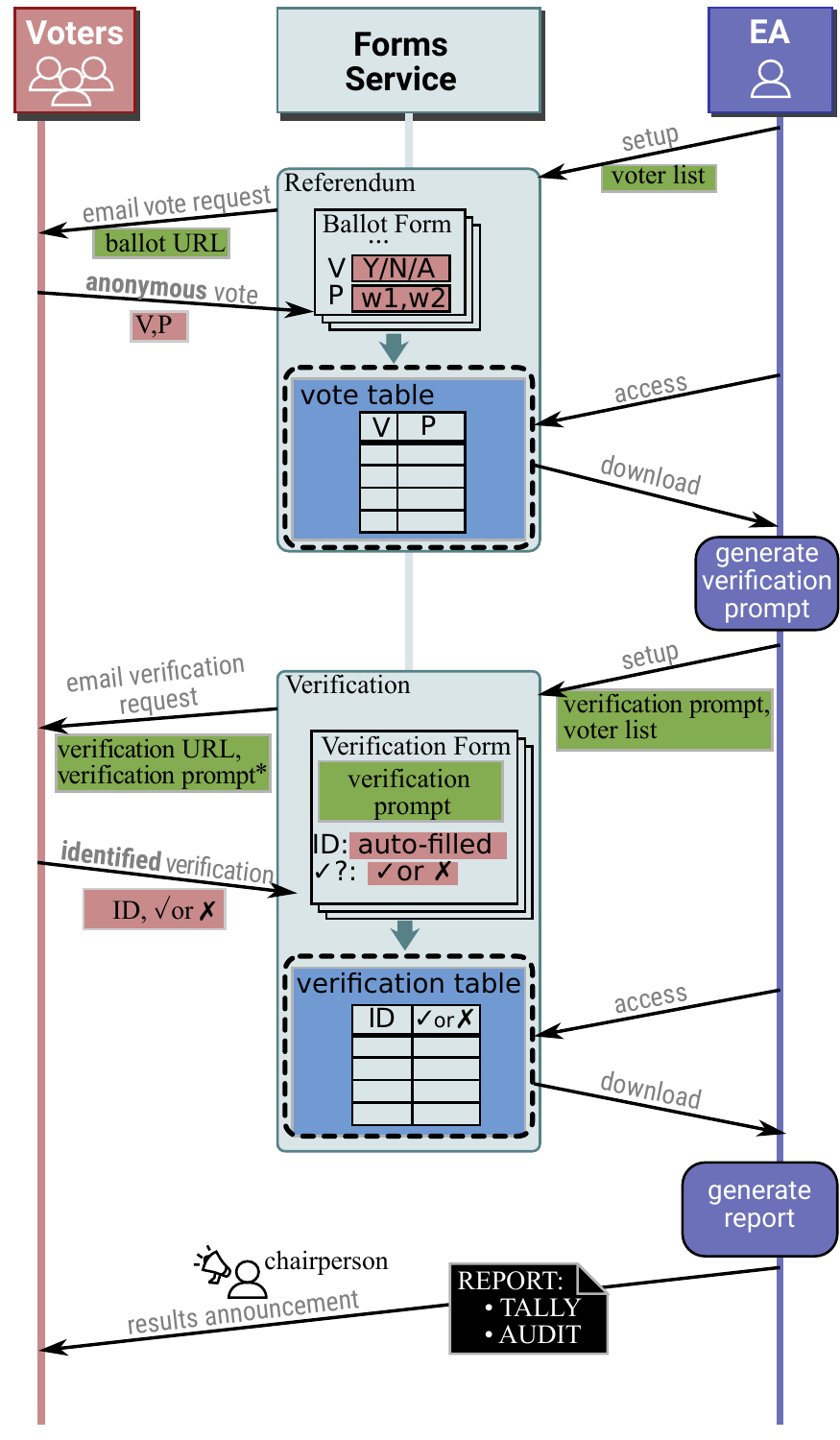}

    \caption{Main steps of Phrase-Verified Voting.
    Each voter enters in a voting form their 
    passphrase and vote $(P,V)$.  After the polls close, each voter verifies the tally and that
    it includes their vote by filling out a verification
    form that includes a verification prompt
    of the $(P,V)$ values. The report includes the tally and all audit data (including the verification prompt and verification table), as explained in Section~\ref{sec:reporting}.}
    
    \label{fig:process}
\end{figure}

%%%%%%%%%%%%%%%%%%%%%%%%%%%%%%%%%%%%%%%%%%%%%%%%%%%%%%%%%%%%%%%%%%%%%%%%%%%

\section{Phase-Verified Voting}
\label{sec:methods}

We describe the \PVV{} system, first presenting an overview, and then separately explaining notation,
in-meeting voting, absentee voting, reporting results, posting audit data, our implementation, dispute resolution, file artifacts, and some helpful voting rules underlying the design.

\subsection{Overview}
\label{sec:overview}

Figure~\ref{fig:process} illustrates the main steps of \PVV{},
and the Appendix includes examples of the ballot and verification forms.

For each referendum, the EA configures a cloud-based selection form with details pertaining to the specific vote. 
Part of this configuration involves restricting access to users within the organization and enabling one-time submission.
The forms service handles authentication of users and enforces a single submission rule, but does not reveal the identity (e.g., email) of voters to the EA. 
It should also have versioning capabilities to provide an audit trail,  to show what was changed and when, if needed.
We used
Google's cloud-based service forms and sheets. 
The EA optionally provides a list of voter emails to facilitate automated distribution of the ballot access link---otherwise the access URL needs to be disseminated through other means.

Each voter anonymously accesses and fills out a form in which they enter their vote $V$ (YES, NO,
ABSTAIN) and a short ``passphrase'' $P$ to enable later anonymous verification. 
The passphrase consists of two words chosen by the voter that are memorable but do not identify the voter. 

The forms service automatically generates a spreadsheet of the $(P,V)$ pairs and a tally of the votes that the EA can access at any time.
The EA may periodically announce the total number of votes cast.
The committee chair announces the close of the election on a video or audio conference call, after which the EA closes the vote by disabling further submissions in the form.

After the poll closes, the EA downloads the vote spreadsheet.
To avoid adding any specialized automation software, the protocol required the EA to format and position all $(P,V)$ pairs using a spreadsheet.
They create the verification prompt by removing timestamps for privacy, sort the data by $V$, separate it into three enumerated lists, and append a tally of $V$.  Figure~\ref{fig:verify} shows an example of the formatted verification prompt.
The innate automatic version control
tracks changes to the tally and allows an audit, in case anyone with access is tempted to make changes to the spreadsheet.

The formatted verification prompt is inserted into a second verification form.
Unlike the ballot form, this second form is configured to  collect voter IDs automatically. As a best practice, the EA distributes
the form access URL by email, embedding a copy of the audit data for voters to keep in their records.  
The forms tool innately automates this process when provided proper form settings and an ID (email) list.

Each voter uses the link to access the verification form to confirm that the verification prompt includes their $(P,V)$ pair and that the tally is computed correctly. 
Because the votes are arranged in numbered lines, each voter can view the accumulation going down the list of votes to confirm the tally without calculation.
Because the passphrases are sorted alphabetically, 
each voter can find theirs easily.
Each voter is required to fill in the verification form, which automatically records their ID.  

Section~\ref{sec:absentee} explains how \PVV{} handles absentee voting.

\subsection{Notation and Terminology} 
\label{sec:notation}

We use the following notation and terminology. 

EA denotes the {\it Election Authority}.
$T_1$ and $T_2$ denote two trusted parties.
The {\it Adjudication Panel} includes $T_1$ and $T_2$ and others.

$V$ denotes a vote (one of YES, NO, ABSTAIN).
$P$ denotes a passphrase, which is a pair of words selected by the voter.
A {\it ballot} is a form that includes a REFERENDUM-ID, referendum date, and place to enter a $(P,V)$ pair.

A {\it referendum} is a specific question about one matter (e.g., the candidate's research work).
Each referendum has a REFERENDUM-ID, which uniquely identifies the referendum.
A {\it committee meeting} is a meeting during which voters make a selection on one or more referenda.
\subsection{In-Meeting Voting}
\label{sec:in-meeting}

The \PVV{} protocol for in-meeting voting works as follows: 

\begin{enumerate}

\item The EA announces all referenda and the dates and times of the committee meetings at which they will take place.

\item For each referendum, the EA announces the number of eligible voters and who they are, by broadcasting and posting a document.
The EA also announces the deadline by which voters may request permission to vote absentee.

\item The EA names the trusted parties $T_1$ and $T_2$, who do not vote in any way in the tenure and promotion process, do not have a stake in the outcome, and who should not be associated with the department.

\item Each voter is expected to select a fresh passphrase $P$ on each voting day. 
The voter may use their $P$ for all referenda held on the same day.

\item For each referendum, the EA starts the voting phase.
People vote by filling out a voting form, in which they enter $P,V$. This form does not record or disclose the voter's identity.
The EA stops the voting phase for the current referendum.

\item The voting form (ballot) creates a table of $(P,V)$ pairs.

\item \label{release}
For each referendum, immediately after voting phase ends,
the EA releases the $(P,V)$ pairs and the vote tally, as described below. 
These voting data are made available to all voters and the Adjudication Panel.

To facilitate verification of the tally, these data are displayed 
in a verification prompt as follows. 
The $(P,V)$ pairs are arranged in
three groups by the three possible votes (YES, NO, ABSTAIN). 
Within each group, the $(P,V)$ pairs are sorted by $P$ values.
Within each group, each $(P,V)$ pair is 
listed on a separate numbered line, with the
first line of each group numbered 1.
See Figure~\ref{fig:verify}.

\item For each referendum, the EA starts the verification phase.
After Step~\ref{release}, each committee member fills out the verification form, which creates a verification table. 
Voters immediately fill out the form during the meeting.
The EA then stops the verification phase for the current referendum.

\item Immediately after the in-meeting verification phase ends, the EA 
broadcasts the verification table to all voters and the Adjudication Panel.

\item Throughout the process, the trusted party $T_2$ receives the tables created by the forms at the same time as does $T_1$, and 
$T_2$ also receives all communications from $T_1$ to the voters. $T_2$ verifies that the audit data 
(see Section~\ref{sec:reporting}) are consistent with the data $T_2$ received from the forms.

% "$T_1$. Trusted party $T_2$" adds redundancy to avoid the awkward "T-1. T_2"(Sentences should not being with math, especially when the previous sentence ends  with math)

\end{enumerate} 

\subsection{Absentee Voting}
\label{sec:absentee}

The university's policy requires that absentee voting be supported and be restricted to voters with legitimate reasons.  Absentee voting introduces challenges to ballot privacy for absentee voters and adds additional administrative complexity.
Each absentee voter may vote at a different time, 
and the entire verification prompt cannot be available 
immediately after absentee voters cast their ballots. \PVV{} handles absentee voting as follows.

\begin{enumerate}
    
\item If an eligible voter cannot attend the meeting, they must send an email to the EA  stating a reason to vote absentee. Permission to vote absentee is granted only for compelling reasons consistent with university policy.

\item To vote absentee, the voter must fill out the voting form at least one hour before the meeting, so that all absentee votes can be included in the
verification form distributed at the meeting.
Immediately after voting absentee, the voter must fill out the {\it absentee acknowledgment form}, which simply records the identity of the voter
and that they voted absentee. 

\item After the final audit data become available, any voter can check that it includes their $(P,V)$ pair. 
Absentee voters do not fill out the verification form---unless they also attend the meeting---because others would be able to identify their vote as one of the few early votes.

\end{enumerate}

\subsection{Reporting Results and Posting Audit Data}
\label{sec:reporting}

During the meeting, and immediately after the verification phase, the EA assembles the audit data, includes them into the report, posts them, and broadcasts them to the committee. These data comprise the REFERENDUM-ID, referendum date, list of eligible voters, list of absentee voters, list of eligible voters who did not vote, vote table, verification prompt, and verification table.
The report should include all data useful for adjudicating a contested election but no information that identifies how anyone voted.

Later, the chair of the promotion committee sends a copy of the committee report to the EA, which attaches the audit data and sends them to all committee members to sign, using 
a document-signing application.

\subsection{Implementation}
\label{sec:implement}

In our implementation, the EA comprised two {designated staff members}, in this case, within the department.
The {staff members} 
generated the ballot form,
formatted the {vote table into a verification prompt to be verified by the voters}, generated the verification form, 
and broadcast these forms to the committee.
The committee chair started and stopped the voting and verification sessions.
In this case, a {designated staff member} of each of the two participating departments served as the second trusted party for the other department.

Voters voted and verified by filling out and submitting a Google form.  
Voters signed the final report using DocuSign\textregistered~\cite{docusign}. %moved before the period as it only applies to one word and not the whole sentence
The EA posted the audit data on the university's 
Box~\cite{box} cloud file system.

To provide some degree of control for eligibility and against proxy voting, the voting form requires that the voter have authenticated themselves via the university's single sign-on system.  
The university's authentication system shares this authentication event with Google, but the voting form does not record the voter's identity.
The verification form also requires this authentication and records the voter's identity.  Recording the voter identity on the verification form helps meet the university requirement to report a list of people who voted, and it enables anyone to detect ineligible or proxy voters.

\subsection{Dispute Resolution}
\label{sec:dispute}

A voter may claim that the verification prompt does not include their
$(P,V)$ pair, but that it contains some $(P,V')$ pair, where $V' \neq V$.
To make such a claim valid, the voter must reveal their name 
and $(P,V)$ pair to the Adjudication Panel. 
The Adjudication panel will broadcast to the voters that someone (without identifying this person) has filed a dispute involving the $(P,V)$ pair. 
For each valid claim, the EA will correct the $V$ in the audit record and correct the tallies accordingly.  

Any other dispute will be adjudicated by the Adjudication Panel, which will deliberate remotely in a conference call. 
The Adjudication Panel should include members who are not associated with the department.
As noted in Section~\ref{sec:security}, \PVV{} cannot resolve a fraudulent claim by a dishonest voter if or when the verification prompt
does not include their $(P,V)$ pair.  

Disputes must be filed within a specified time, say 48 hours, from the time the EA posts the audit data.
The Adjudication Panel will report the number of claims it received and, for each, a summary of the general nature of the claim and whether and how it was resolved. There were no uses of the adjudication protocol in our elections.

%%%%%%%%%%%%%%%%%%%%%%%%%%%%%%%%%%%%%%%%%%%%%%%%%%%%%%%%%%%%%%%%%
\subsection{File Artifacts}
\label{sec:artifacts}

{The key surviving document from the process is the final report; 
special handling should be considered for
a few other resulting intermediate documents. 
Two such documents are the original downloaded vote table and the downloaded verification table, typically downloaded in a spreadsheet format.  When using the Google Forms service, these spreadsheets include timestamps for each vote. These originals also contain the results of votes. }

{The other surviving documents would be in the the email verification request: we choose to include the verification prompt in the emailed verification request, though this inclusion is optional 
(denoted verification prompt* in Figure~\ref{fig:process}). With the Google Forms service, this action
was done by placing the verification prompt in the form description.  
By comparing verification records, users could detect changes 
to the verification prompt between users.
Providing this information in the emailed request could discourage users from verifying the information directly in the online form.  If the results of the vote are highly sensitive and email control policies are not appropriate, the verification prompt
should not be included in the email verification request.  Finally, the forms service preserves a version of the download records, so destruction of these 
documents after the opportunity for dispute resolution
should be considered.}
%%%%%%%%%%%%%%%%%%%%%%%%%%%%%%%%%%%%%%%%%%%%%%%%%%%%%%

\subsection{Rules for Voting} 
\label{sec:rules}

We state six rules for voting and operational policies that enhance the voting process: 

\begin{enumerate}

    \item All voters, including absentee voters to the extent possible, must use the same voting procedure (i.e., fill out an identical but individual form privately).
    
    \item All absentee votes are due at least one hour before the scheduled committee meeting, at which a referendum is to take place. Each absentee voter must acknowledge their vote immediately after voting.
    
    \item \label{late} Late votes are not permitted. The voting form will not accept any votes after the close of voting.
    
    \item Election outcomes are announced at the committee meeting, immediately after voting ends, by displaying the tables from the cloud. No preliminary results are announced before voting ends.
    
    \item \label{required} Each in-meeting voter is required to vote and to engage in a vote verification step. Each voter present at the meeting must verify their vote during the meeting. Each referendum is verified separately.
    
    \item Later in the semester, using a document-signing application, each committee member signs the committee report, which includes the two tables produced by the voting and verification steps.
    
\end{enumerate}

Rule~\ref{late} helps ensure that the voting process will terminate during the meeting and in a way that can be verified by the voters during the meeting.

Rule~\ref{required} enhances outcome integrity and facilitates the creation of a list of eligible voters who voted, and a list of eligible voters who did not vote, as required by university policy.

%%%%%%%%%%%%%%%%%%%%%%%%%%%%%%%%%%%%%%%%%%%%%%%%%%%%%%%%%%%%%%%%%%%%
\section{Security Notes} 
\label{sec:security}

% somewhere discuss trust in google

This section details the system's security properties with regard to outcome integrity, dispute resolution, discreditation attacks, ballot privacy,
voter identification, and logging.

Importantly, the system is ``software independent,''~\cite{rivest2008notion} in that any software fault that affects the tally can be noticed by voters. 

\subsection{Outcome Integrity}
\label{sec:integrity}

The system provides strong outcome integrity because voters have opportunities to detect any manipulation of the tally.  
Since they know the number of eligible voters and the number of people who voted, voters can detect if
additional votes were inserted into the tally by inspecting the vote counts and by noticing 
too many $(P,V)$ entries in the verification form. 

This opportunity to detect manipulation of the tally assumes that the EA provides accurate data, especially the list of voters, in a timely fashion. Any anomalous behavior by the EA is detectable in the audit.

Each voter can observe that the verification form lists their $(P,V)$ pair.
If the $P$ values are distinct, each voter can detect any attack that changed their $(P,V)$ pair.
If exactly two voters pick the same $P$ value, the only case where an adversary could change one vote without detection is when both voters also cast the same corresponding vote. 
Absentee voters can verify their votes from the final audit data.

It is valuable that all voters  participate in the verification process.
If the adversary knew that a particular voter would not verify, and could identify the corresponding $(P,V)$ pair, they could attempt to modify that pair without detection.

If the adversary compromises the voter's machine, or
mounts a man-in-the-middle attack, the
adversary could attempt to modify a vote and display a compromised verification form to the corresponding voter.  This attack risks detection if the voter checks the verification form or final audit data from a separate, uncompromised machine.

\subsection{Dispute Resolution and Discreditation Attacks}
\label{sec:security-dispute}

A voter might claim that their $(P,V)$ pair is
not present in the verification form.
When only the $V$ value is in dispute, the dispute resolution procedure provides an adequate resolution, albeit at the cost of the voter revealing their identity to the Adjudication Panel. Specifically, the Adjudication Panel can adjust the tally by replacing the disputed $(P,V)$ pair with the corrected one offered by the complainant and posting this action into the audit record
(see Section~\ref{sec:dispute}).

If the adversary changes one or more $P$ values, voters can detect such change, regardless of whether the adversary additionally modifies any $V$ values.
The system does not have capability to correct such anomalies. 
Relatedly, the system does not prevent a dishonest voter from claiming falsely that  the verification form does not list their $P$ value. Therefore, the system is vulnerable to a discreditation attack by dishonest voters who falsely claim their $P$ is not present. Adversaries who modify one or more $P$ values can be detected by the logs of changes to the forms (see Section~\ref{sec:logging}).

\subsection{Ballot Privacy}
\label{sec:privacy}

The system offers reasonable privacy to honest voters, but none to dishonest voters.
A dishonest voter can intentionally identify themself through their choice of $P$ value, similar to the way a dishonest voter could write an identifying mark on a paper ballot. 
Unfortunately, this limitation facilitates undue influence via vote buying and coercion. 
For example, a coercer could demand that the victim vote with a particular $(P,V)$ pair.
Mitigating undue influence is the most daunting challenge of Internet voting~\cite{votexx}.

Just because a $P$ value might be someone's name or initials does not necessarily mean that the voter is the identified person.  
It is possible that a dishonest voter intentionally entered the $P$ value in question for the purpose of making it appear that the identified person voted in a certain way.

An adversary who compromises a voter's machine or monitors communication traffic could learn how the voter voted.
Furthermore, although the voting form does not overtly identify the voter, it is possible that careful technical analysis of network traffic could identify the voter from an intercepted voting form.

A corrupt EA (e.g., corrupt $T_1$) could attempt the following attack: send a custom voting form to each voter, resulting in each voter creating a separate vote table. This attack could be detected by $T_2$ and by anyone
inspecting the final audit data.  The vulnerability highlights the importance of the voting form URL sent to voters.

There are fundamental limits on ballot privacy resulting from the number and identity of regular and absentee voters.  
This limitation particularly affects absentee voters because there are typically few of them. 
Although the  $(P,V)$ pairs of the absentee voters
are not identified to the entire electorate as such, these votes are automatically entered into a cloud-based spreadsheet before the regular votes are entered.
Therefore, anyone with access to this spreadsheet, including the EA, would likely be able to identify the absentee $(P,V)$ pairs.  
If, for example, there were two absentee voters, and they both voted NO, then anyone who knows the absentee votes would know how each absentee voter voted.

\subsection{Voter Identification}
\label{sec:identification}

Because the verification form (but not the voting form) requires the voter to be signed into an account, the system provides some degree of assurance and identification of who filled out the verification form.

\subsection{Logging}
\label{sec:logging}

There are several points in the process where the
system creates an indelible record and commitment as a result of sending emails and posting files on a cloud-based file system with automatic versioning records.
In particular, the EA sends voting and verification forms by email and posts the final audit data on a cloud-based file system.

%%%%%%%%%%%%%%%%%%%%%%%%%%%%%%%%%%%%%%%%%%%%%%%%%%%%%%%%%%%%%%%%%%%%%%%%%%%
 \section{Evaluation and User Study}
 \label{sec:eval}
 
 From October 6 to November 23, 2020, two departments, Computer Science and Electrical Engineering (CSEE) and Information Systems (IS), used the \PVV{} system, conducting a total of 26 binding referenda on promotion cases, each involving 8 to 26 voters.
 Subsequently, 18 of 43 distinct voters answered a  follow-up survey about their experiences.  One EA member for CSEE also answered the survey.
 
\subsection{Survey}
\label{sec:survey}

The survey comprised five parts.  Parts~1--3
explored user reactions on a five-point Likert scale.
Part~4 solicited free-form responses, and Part~5 
solicited demographic information.
Part~1 had three questions comparing \PVV{} to vote-by-mail on matters of usability, trust, and privacy.
Part~2 had three questions
comparing \PVV{} to the paper ballot system used in the previous year on matters of usability, trust, and privacy.
Part~3 had two questions on whether the verification step increased their trust in the system and whether it was worth the additional cost. 

\begin{figure}[H]
    \centering
    \includegraphics[width=0.88\columnwidth]{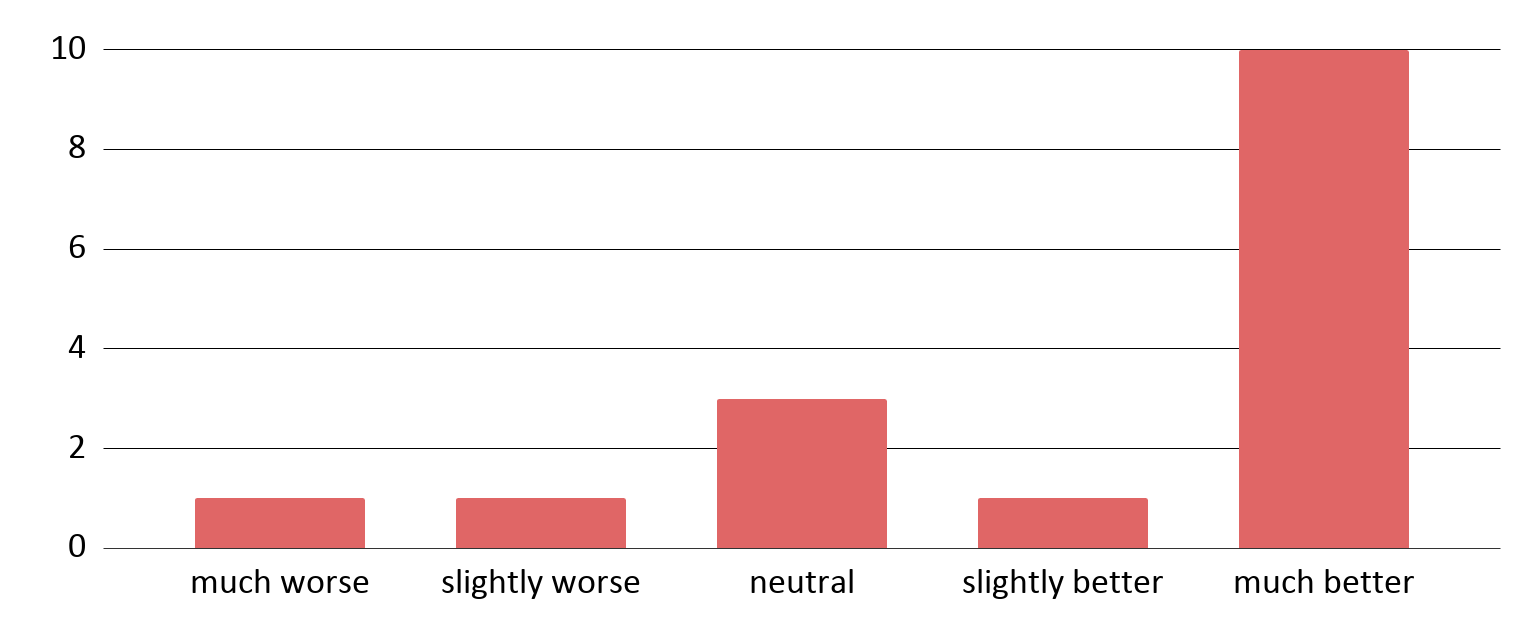}
    \caption{Answers to ``Comparing the online system versus emailing your vote: How easy was it to vote?''}
    \label{fig:likert-results-2-1}
\end{figure}

\begin{figure}[H]
    \centering
    \includegraphics[width=0.88\columnwidth]{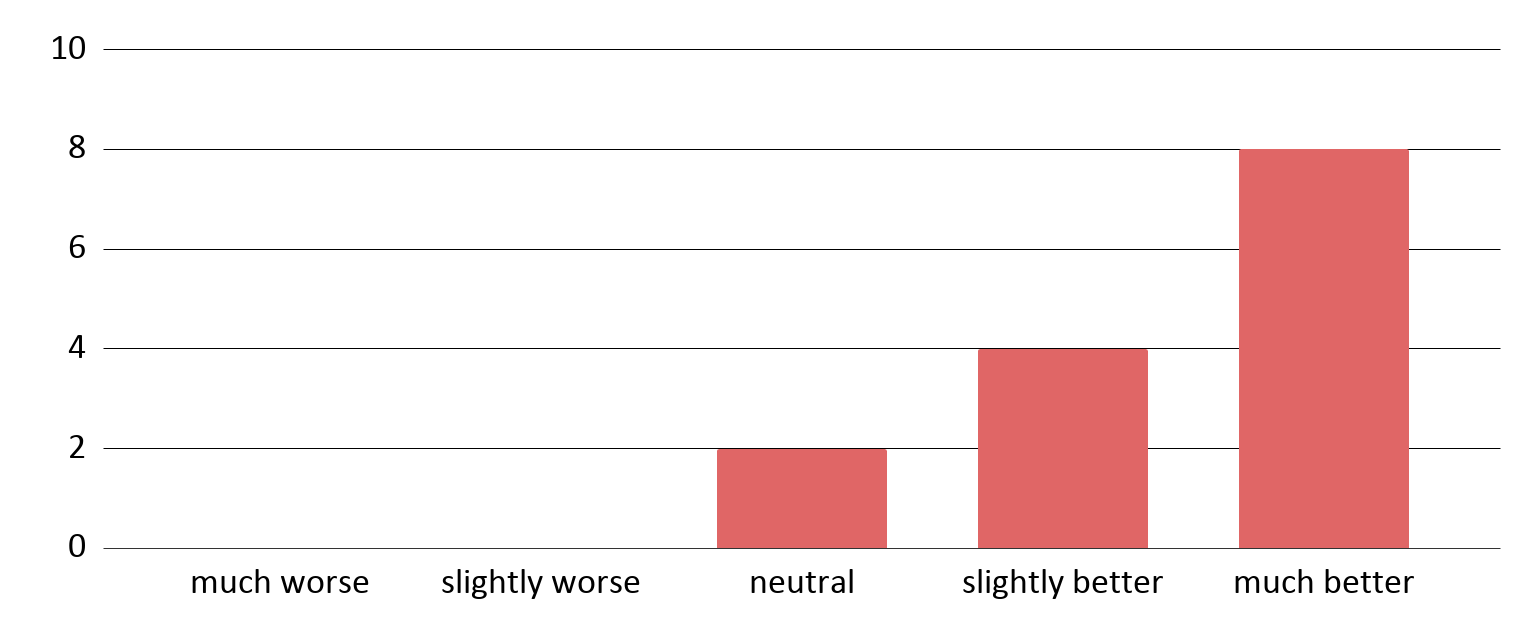}
    \caption{Answers to ``Comparing the online system versus emailing your vote: How much do you trust the result?''}
    \label{fig:likert-results-2-2}
\end{figure}

\begin{figure}[H]
    \centering
    \includegraphics[width=0.88\columnwidth]{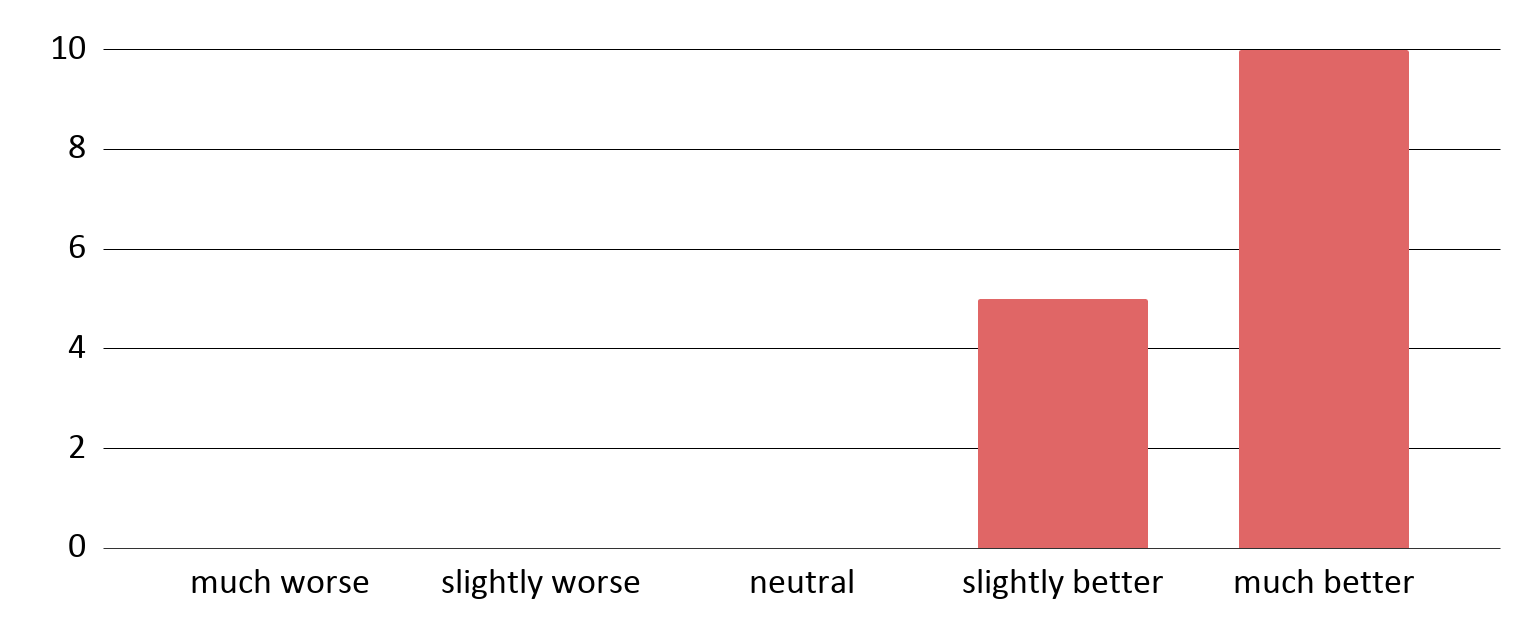}
    \caption{Answers to ``Comparing the online system versus emailing your vote: How sure are you that your vote is private?''}
    \label{fig:likert-results-2-3}
\end{figure}

\begin{figure}[H]
    \centering
    \includegraphics[width=0.88\columnwidth]{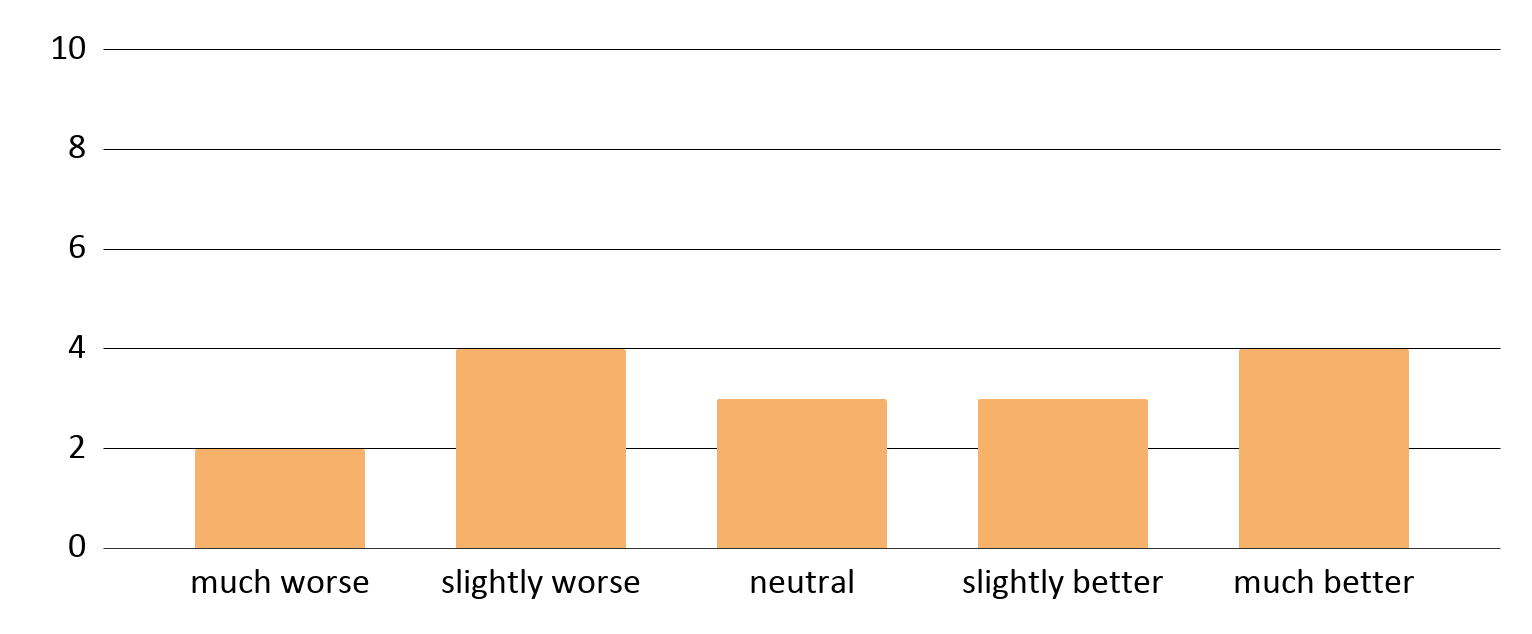}
    \caption{Answers to ``Comparing the online system versus in-person paper ballot voting system you used last year: How easy was it to vote?''}
    \label{fig:likert-results-2-4}
\end{figure}

\begin{figure}[H]
    \centering
    \includegraphics[width=0.88\columnwidth]{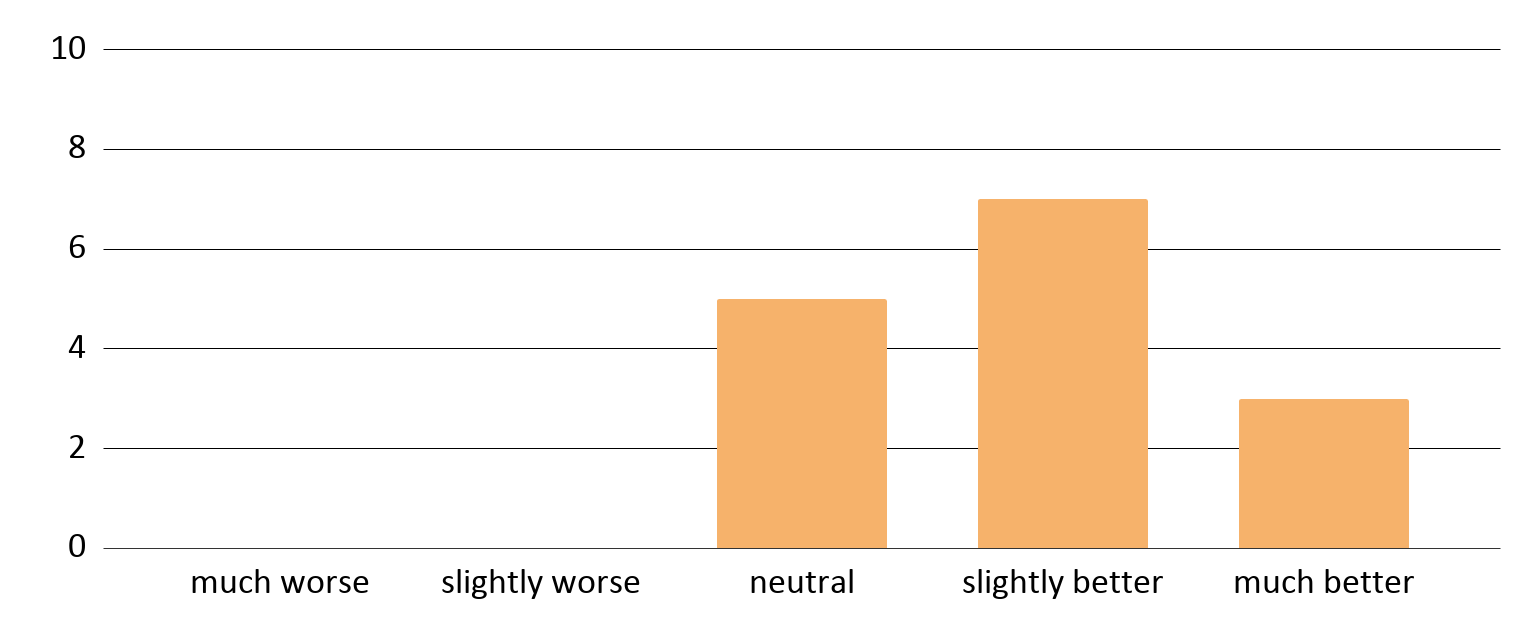}
    \caption{Answers to ``Comparing the online system versus in-person paper ballot voting system you used last year: How much do you trust the result?''}
    \label{fig:likert-results-2-5}
\end{figure}

\begin{figure}[H]
    \centering
    \includegraphics[width=0.88\columnwidth]{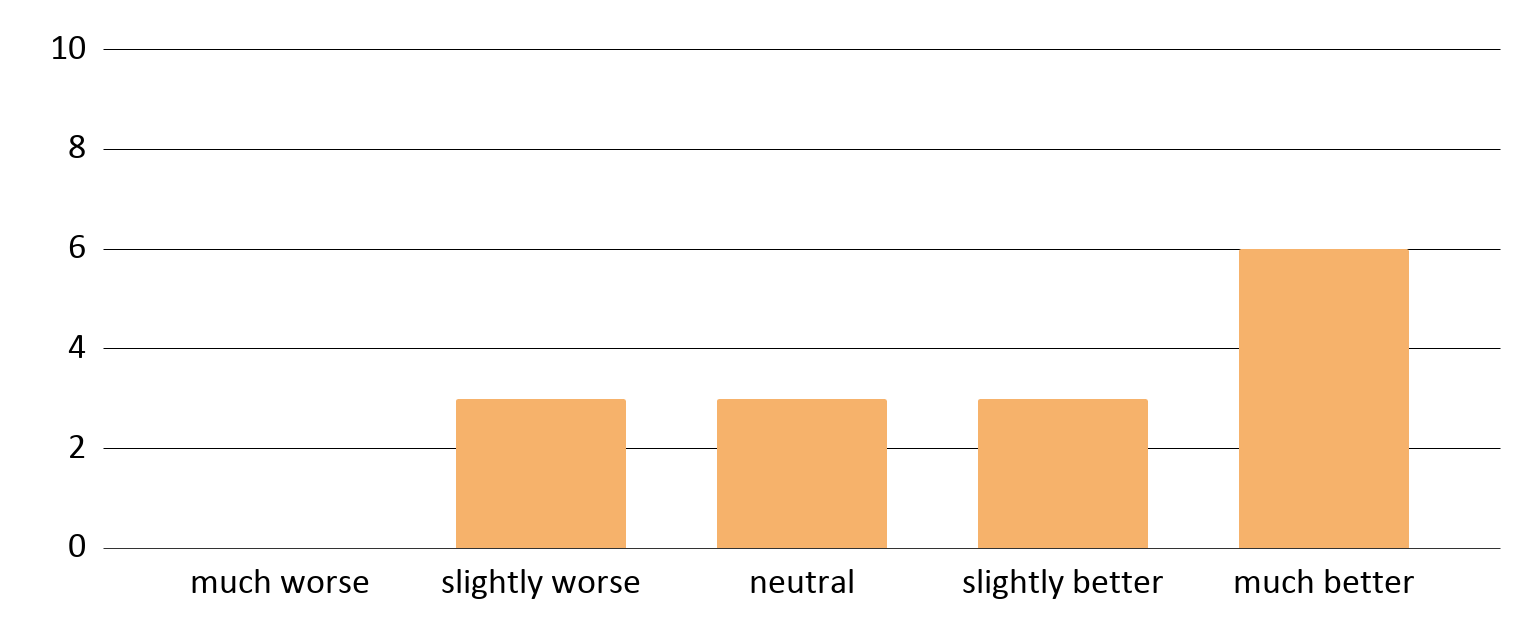}
    \caption{Answers to ``Comparing the online system versus in-person paper ballot voting system you used last year: How sure are you that your vote is private?''}
    \label{fig:likert-results-2-6}
\end{figure}

\begin{figure}[H]
    \centering
    \includegraphics[width=0.88\columnwidth]{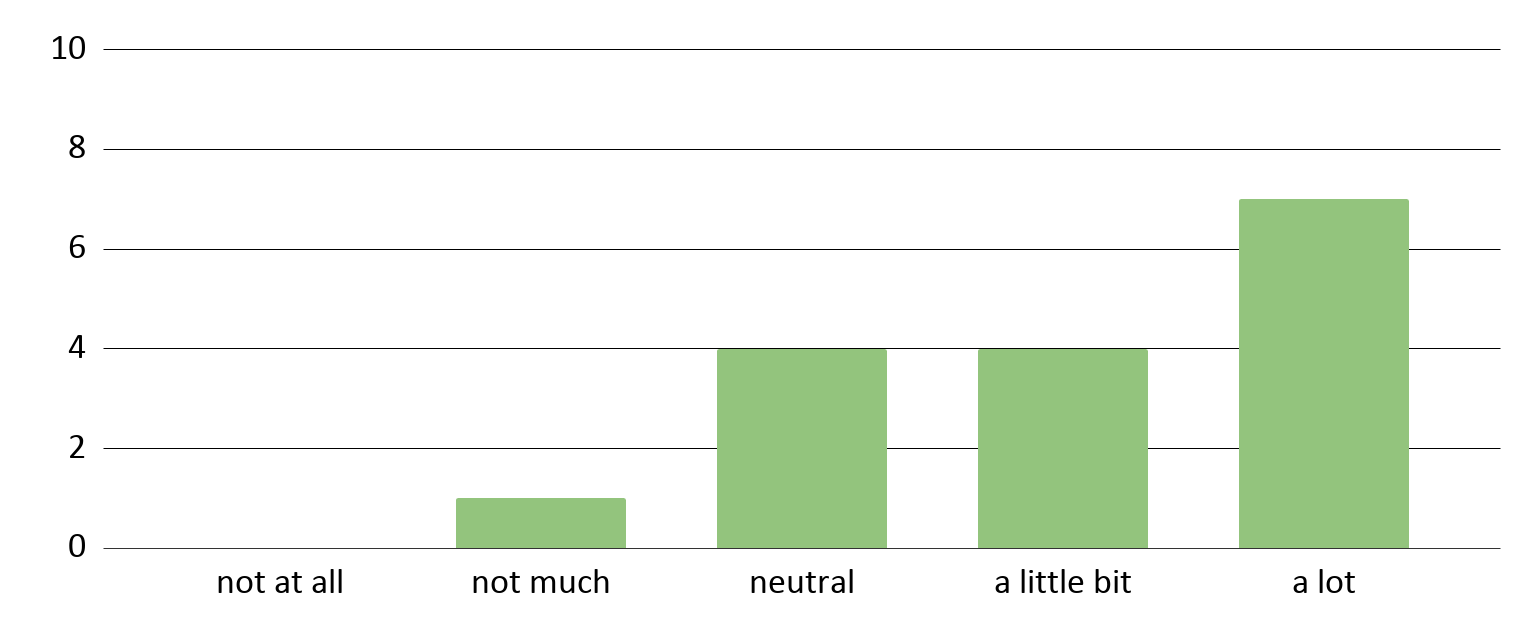}
    \caption{Answers to ``Did the verification step increase your confidence in the result?''}
    \label{fig:likert-results-2-7}
\end{figure}

\begin{figure}[H]
    \centering
    \includegraphics[width=0.88\columnwidth]{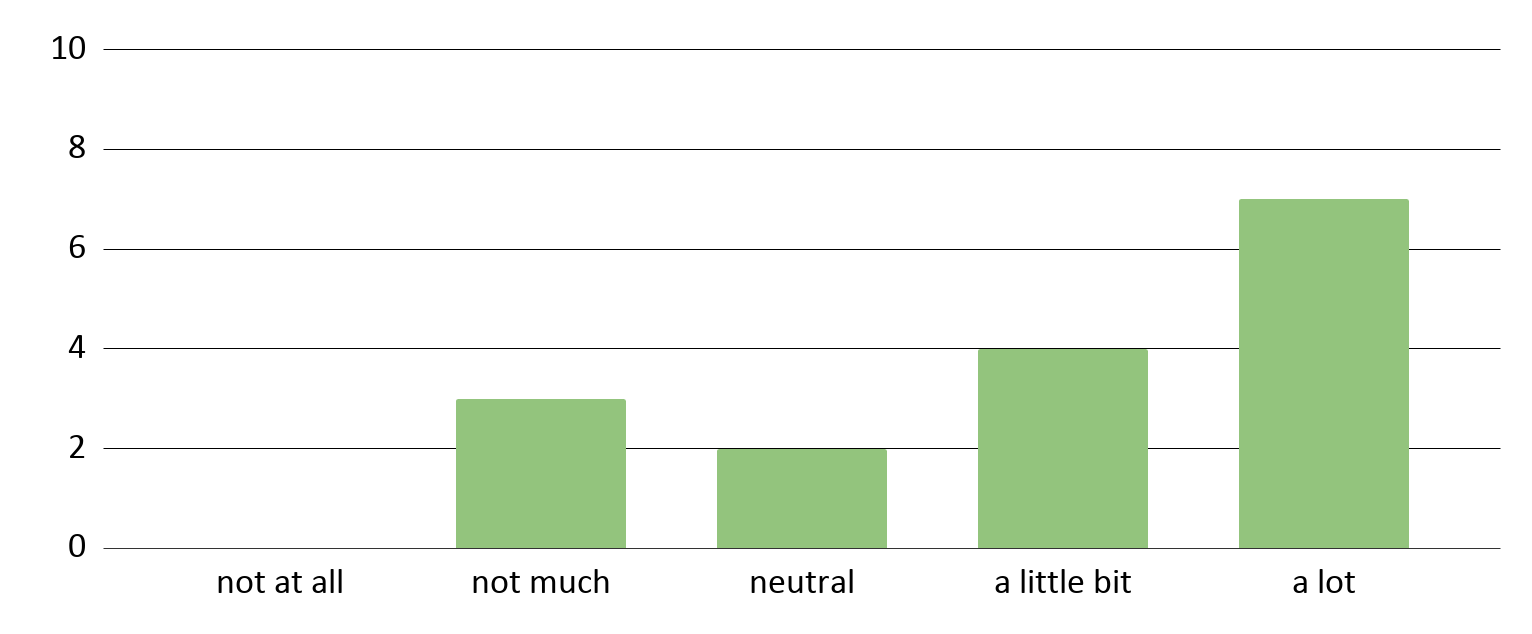}
    \caption{Answers to ``Is the verification step worth the additional effort?''}
    \label{fig:likert-results-2-8}
\end{figure}

From one question in Part~5, we learned that none of the respondents had any experience with verifiable voting systems. 
Because responses from IS and CSEE were substantially similar, we present histograms (Figures~\ref{fig:likert-results-2-1} to \ref{fig:likert-results-2-8}) show the combined responses of the 5 respondents from IS and 12 from CSEE. 
They exclude the single administrator's responses.

Assigning numerical grades from 1 to 5 to each of the Likert items shows that voters found \PVV{} trustworthy (mean 4.47) and private (4.47), and a bit easier (4.00) than email voting. 
Only two people rated it worse, and only in terms of usability (due to the verification step). 
The preference is less marked against in-person voting but still present. One voter from each department consistently gave the highest grade on all questions, but all other respondents changed their grades question-by-question, showing that they probably gave each question some thought. 

\subsection{Open-Ended Feedback}
\label{sec:feedback}

In open-ended feedback from voters, seven respondents considered the delay between voting and verification too long (see Section~\ref{sec:recs}).
Some voters appreciated the convenience of our system and noted that, after an initial learning curve, it was faster than using paper ballots. 

Several voters stated that allowing people to verify their vote and the tally enhanced trust.
Some also appreciated that it allowed them to vote at a time of their convenience (through absentee voting) and that it could engage people who could not come to in-person meetings but who could attend virtually. One voter wrote that, despite the system being cumbersome, it was worth the cost to give people a more hands-on experience with voting issues of which everyone should be aware. Finally, one indicated that they would prefer to verify just their own vote and to have a single voter verify the tally.

The administrator liked the paper reduction but did not like having to remind voters to perform steps in a timely manner. They remarked that it was tedious from an administrative standpoint to move the data around, but once in place, ``the process was in sync and it seemed to work well.''

\section{Discussion}
\label{sec:discussion}

We now discuss our results and several issues that arose during our study.
These issues include our major design decisions, usability, role and trust of the EA, mandatory voting and verification, choice of passphrases, limitations, recommendations, and open problems.

%E : I don't know how to integrate the next two nicely, but they should probably go somewhere
% E: possibility of having the verification after the meeting (and all the problems it creates, including the fact that the people who want to have it after might simply want not to be accountable for the fact they didn't verify)... 

% E: peculiarity of t&p voting with often lopsided votes and where having the majority isn't the only thing that matters

\subsection{Discussion of Results} %%%%%%%%%%%%%%%
\label{sec:results-discussion}

Typically voters do not like or trust new voting system~\cite{Vassil2016diffusioninternetvoting.}.
When they try a new one, they usually take longer to use it and make more mistakes. However, first-time users of the \PVV{} system preferred it to the paper ballot system they had used the previous year.  It is likely that voters accepted \PVV{} in part because it used the familiar interface of Google Forms.  

%[Careful: I do not think we have data to support eh claims about speed and mistakes. They had more confidence in its privacy and outcome integrity.  

As \PVV{} adds a verification step, it is not surprising that some voters found it more cumbersome. In the open-ended comments, the most significant problem mentioned was the five to ten minute delay between voting and verification (see Section~\ref{sec:recs}). 
Still, it is notable\footnote{Voter interest in verification might have been affected by the political climate in the USA at the time, with a lively debate on the security of voting systems.} that voters felt strongly that the verification step increased their confidence in the results and was worth the extra effort. 

\subsection{Major Design Decisions} %%%%%%%%%%%%%%%
\label{sec:decisions}

We made the following key design decisions.
\begin{enumerate}

 \item The outcome integrity does not depend on any software or individual operating correctly, making it software independent.

 \end{enumerate}

\pagebreak
%%%%%%%%%%%%%%%%%%%%%%%%%%%%%%%%%%%%%%%%%%%%%%%%%%%%%%%%%%%%%%%%%%%%%%%%%%
%\subsection{Likert Results} 
%\label{sec:likert}
% the two prose pars are redundant with the section text, and they
%      complicate fig placement
%Figures~\ref{fig:likert-results-2-1}--\ref{fig:likert-results-2-8} present the main survey results. Figures~\ref{fig:likert-results-2-1}--\ref{fig:likert-results-2-3} show the response histograms for the comparison of \PVV{} with emailing votes.

% Figures~\ref{fig:likert-results-2-4} through \ref{fig:likert-results-2-6} show the comparison of \PVV{} with the simple paper ballot voting system used the previous year. Figures~\ref{fig:likert-results-2-7}--\ref{fig:likert-results-2-8} show the response histograms for the questions about verification. 

\clearpage
%%%%%%%%%%%%%%%%%%%%%%%%%%%%%%%%%%%%%%%%%%%%%%%%%%%%%%%%%%%%%%%%%%%%%%%%%%%

\begin{enumerate}
\setcounter{enumi}{1}

 \item The design uses standard tools (especially forms and spreadsheets) familiar to voters and system administrators. This decision contributed to the usability and implementation simplicity of our system.

 \item The verification form organizes the verification table to help voters  find their passphrase and verify the tally without any complex counting (see Section~\ref{sec:usability}).

 \item The system does not use any complex protocols or cryptography.  This decision contributed to the usability while still providing an appropriately lightweight level of privacy and integrity for the application.

 \item The system did not accept any late votes; absentee votes had to be cast prior to the meeting; and in-meeting voters verified their votes during the meeting.  Although some voters may have preferred the ability to skip the verification step, this decision avoided many difficult issues experienced in the past and enabled the EA to announce the results with greater assurance during the meeting.

 \item The system treats absentee voting as closely as possible to in-meeting voting. Although the special demands of absentee voting created some difficulties, this decision helped  minimize the complexity required to deal with absentee voting.
 
 \item The system sends the URL to voters asynchronously via email. This decision helped voters with limited Internet connections to participate in the meeting. To mitigate the custom voting form attacked mentioned in
 Section~\ref{sec:privacy}, it is helpful to post the voting form URL in the meeting and to include it in the audit data.
 
 \item For simplicity, we designed a system that does not protect against certain attacks, such as discreditation attacks.  We did so knowing that the stakeholders could adopt a policy that,
 if such attacks were detected, a second referendum could be held using a more secure and more complex system (which would cause a delay). Because boardroom referenda are generally easier to organize than large-scale elections, this delay would be limited. Moreover, an attacker wanting to delay the referendum could do so at little cost by claiming to have issues with their computer.
 
 \end{enumerate}

\subsection{Role and Trust of the EA} %%%%%%%%%%%%%%%
\label{sec:EA}

The Election Administration performed the following functions, communicating only via computer:
(1)~Produced the list of referenda and their eligible voters.
(2)~Identified the absentee voters.
(3)~Generated the voting forms.
(4)~Generated the verification forms.
(5)~Started and stopped the voting and verification phases of each referendum.
(6)~Archived the audit data.
(7)~Distributed and collected the final report to be signed by all voters.
(8)~Received out-of-band communications for exceptional circumstances, e.g., when one voter's Internet connection failed during a meeting.
    
%\begin{enumerate} % took out to save space
%\end{enumerate}

In the university promotion referenda, the EA was implemented with two people
who happened to be staff members of the department.
Opportunities for fraud could be reduced by requiring the EA not to be  associated with the department or people in it.
Although not part of the EA, the chair of the promotion committee attended each meeting and started and stopped the voting and verification phases. 

% academic affairs manager, 

None of the functions requires any trust in the EA for outcome integrity.  Any attempt by the EA to alter votes could be detected by the voters.  Similarly, except for the out-of-band communications, none of the functions requires any trust in the EA for ballot privacy.
The EA, however, could easily cause a denial of service, for example by failing to produce the required forms, claiming that the computer
crashed, or intentionally introducing errors into the tables.

The EA has a strong position of access to crucial documents, including the voting and verification forms, that could facilitate attacks on privacy.
For example, a malicious form might have hidden functionality that spied on the voter.  
Any adversary, however, could attempt to intercept
and replace the legitimate forms.  
Similarly, any adversary could intercept and inspect completed forms for possible metadata that might be helpful in privacy attacks. 

\subsection{Usability} %%%%%%%%%%%%%%%
\label{sec:usability}

Several details of the system enhance its usability.  
First, selecting ballot choices in a form is a familiar task for everyone because they have filled out other surveys with this tool. Second, the verification prompt facilitates checking by grouping similar votes (YES, NO, ABSTAIN) and numbering the lines, helping voters verify the tallies without having to count. Ordering the passphrases alphabetically reduces effort for voters to find their passphrase in the list.
Third, voters can remember passphrases more easily than codenumbers (see Section~\ref{sec:passphrases}).

The \PVV{} experiment revealed three problems.  
First, despite careful instructions, several voters initially did not understand the concept of verification and what actions they had to carry out.
After one or two referenda, they became comfortable with the process.
Second, despite holding a PhD in computer science and being familiar with forms, several voters forgot to select ``submit'' after filling out a form. 
Third, there was a significant delay (about five to ten minutes) between the end of voting and verification caused by the manual creation of the verification table (see Section~\ref{sec:recs}).
No complaints presented  the opportunity to test the adjudication process.

\PVV{} is intended for use in small private elections; it would likely be difficult to use for larger elections (e.g., much greater than, say, 100 voters).

\subsection{Mandatory Voting and Verification} %%%%%%%%%%%%%%%
\label{sec:mandatory}

The referenda were conducted with mandatory voting and verification.  By university policy, all eligible voters not on sabbatical were required to vote.  Faculty on sabbatical leave were allowed, but not required, to vote. 

Members of the CSEE promotion committee voted to mandate verification, 
as did the IS committee.  
The reason was to enhance outcome integrity.
An adversary might attempt to modify the votes of anyone who would likely not verify.
Although some voters showed impatience with the extra verification step, all voters tolerated and succeeded at performing it.

Furthermore, university policy requires that the promotion committee report identify who voted, how many people voted absentee, and who did not vote (with explanation).  
Although such tabulations might seem simple to perform accurately, practice shows otherwise.  
In the past, there had been minor discrepancies caused by confusion during meetings, absentee voting, late voting, multiple referenda, and the fact that different sets of people were eligible to vote on different referenda.  
\PVV{} and its associated policies (e.g., no late voting) facilitated the accurate tabulation of these counts.

\subsection{Passphrase Choice} %%%%%%%%%%%%%%%
\label{sec:passphrases}

The design of \PVV{} considered a few possible variations for a unique identifying code before settling on two-word passphrases. This choice leverages people's ability to remember word pairs better than numbers~\cite{Miller1956magicalnumberseven}.
Based on previous experiments on word choice, we estimate that there is less than 0.1\% chance that two voters in a small group would select the same passphrase~\cite{Blanchard2018Improvingsecurityusability,Bonneau2012Linguisticpropertiesmulti}.

A major issue with voter-chosen passphrases is that such passphrases enable coercion: a coercer could demand that the victim choose a particular passphrase.  To mitigate this concern, one might consider restricting the choice of passphrases to a given block or cloud of words, which would enforce a low probability of collision and high memorability~\cite{Blanchard2018Improvingsecurityusability}, and could be done by adding a software layer on the forms service. 
It could also be achieved---except for absentee voters---by 
displaying the list of words during the meeting.
In both cases, the approach could be compromised if
the coercer had access to the cloud of words seen by the voter, and might demand that the voter select the words in the upper-left and upper-right corners of the block or cloud.

Another approach is to use random word generators, 
such as~\cite{rw1} or~\cite{rw2},
which would dramatically reduce the chance that two voters choose the same passphrase. We recommended using one (and indicated the two above), but did not mandate this practice to avoid introducing software that might violate ballot privacy.

% moved to bib to save space
%\url{https://randomwordgenerator.com/} or %\url{https://www.wordgenerator.net/random-wordgenerator.php}, 

Passphrases chosen in our study and an analysis  revealed that some voters did not respect the two-word rule or chose low-entropy passphrases, including ``abc def''  or ``dog cat''. Moreover, in at least one referenda, the first word of one passphrase was the three-letter initials of a committee member.
It remains a mystery whether this voter was actually the identified committee member, or whether some other voter had intentionally created the
false impression of being that member.
This situation could be problematic but the freedom to choose one's passphrase means that it is difficult to distinguish a voter having fun by writing weird passphrases from someone trying to be identifiable. As such, the guessable link between a passphrase and its potential author is poorly actionable information. 

Having multiple referenda at the same meeting raised additional usability issues with passphrases.  Because it would be difficult to remember multiple passphrases, it may seem reasonable for voters to reuse their passphrases for different referenda conducted on the same day.
Doing so creates two issues.  
First, it enables anyone to link the votes from the different referenda.  
Second, a malicious voter could intentionally reuse some other voter's passphrase for the purpose of creating confusion; such behavior, however, is unlikely to enable a vote-changing attack.
To limit the number of instructions, we chose to let voters do as they wanted, with a minority changing their passphrases between votes.

\subsection{Limitations of the User Study} %%%%%%%%%%%%%%%
\label{sec:limits}

Limitations of the study include a small voter population and possible bias in the survey respondents. Ideally, in user studies, the respondents should not know the creators of the systems tested to avoid being biased. Some of the voters (some of the anonymous respondents) knew two of the creators, but many of them did not, including respondents from IS.

\subsection{Reducing the Verification Delay} %%%%%%%%%%%%%%%
\label{sec:recs}

A main recommendation is to reduce the delay 
in sending out the verification form caused by the manual construction of the verification prompt.  During each referendum, after the close of voting, it took the EA approximately five to ten minutes to send out the verification form.  The main cause of this delay was constructing the verification prompt. 
Although this delay might be less than that to count paper ballots, it tested voter patience, and with multiple referenda held during the same one-hour meeting, it risked extending the election past the allotted time. It would be better to generate
the verification table automatically.

Automatically constructing the verification prompt is a simple task algorithmically. The main challenge is doing so in a way that does not compromise privacy, integrity, and helps voters see the results easily. 
On the positive side, because voters could detect errors in the
verification prompt, generating it automatically does not threaten
the software independence of \PVV{}.
On the negative side, running any software on the EA's computer risks compromise of that machine, which could threaten denial-of-service and
to some extent erosion of ballot privacy.
Furthermore, introducing software into the voting process might cause conflict with institutional policies that seek clarity and transparency of the voting process.
After user feedback, we implemented a method to instantly generate the verification prompt using a second Google sheet, with the first sheet having the automatic verification table.  However, the system as described is transparent. Automatically generating the verification prompt should be introduced only after voters (and other parties) are comfortable with the process, since it introduces multiple spreadsheet functions. Alternatively, one could create a remote cloud service to accept a vote table and return a verification prompt, rather than running custom scripts on the EA's computer.

%\footnote{This helpful functionality could be used by many organizations and could be vetted and periodically randomly tested by anyone.  The service would have no need to know the identity of the user or referendum. The service could authenticate itself and include a cryptographic hash of the input vote table in its output. Using such a service would be convenient and introduce less risk to the EA's computer than running software locally.} 

% Be aware that some journals have a policy to forbid most footnotes.

\subsection{Open Problems} %%%%%%%%%%%%%%%
\label{sec:open}

In comparison with the existing options of Helios, voice vote, and emailing plaintext votes, the \PVV{} system offers a new balance between  outcome integrity, ballot privacy, usability, and ease of system configuration.  New  systems could be developed to offer different combinations of these properties better suited to different scenarios. 

\PVV{} itself could also be adapted to other voting scenarios with few voters, and one could test its efficacy in those environments, especially in a comparative user study between \PVV{} and other systems. 
It would be interesting to explore how \PVV{} could be used as a teaching tool to help people learn about 
verified voting. This exploration
could be done by inquiring about their knowledge, understanding, and opinions both before and after using the system. To date, \PVV{} has been tested only as an emergency replacement to improve security. 

Finally, an analysis of the system using formal methods would also be welcome toward the goal of establishing better security guarantees. 

%%%%%%%%%%%%%%%%%%%%%%%%%%%%%%%%%%%%%%%%%%%%%%%%%%%%%%%%%%%%%%%%%%%%%%%%%%%
\section{Conclusion}
\label{sec:conclusion}

We have proposed and deployed a transparent system  for private remote boardroom voting that is low-tech, voter-verifiable. \PVV{} offers a new option that emphasizes a simple setup and deployment and few steps for the user, while still providing voters the ability to verify that their vote was correctly recorded and tabulated. It demonstrates how simplified versions of modern voting security approaches can be used in new and different kinds of elections. \PVV{} is well-suited for small (say, less than 100 people) groups with some degree of trust when voting on low-stakes referenda.

\PVV{} demonstrates a few-minute online referendum process in which every voter assures that their vote is included in a final tally, without requiring special software or difficult-to-arrange systems. It offers greater ballot privacy and outcome integrity than does the commonly used alternative of sending votes to a trusted party by email. The system, however, does
not provide as much privacy and integrity as does the more complex to set up
Helios system. \PVV{} gently introduces voters to the
idea of voter-verified systems.
It can also give people an understanding of verification and the motivation to consider adopting verifiable voting systems, which has been an issue in the past~\cite{Karayumak2011Userstudyimproved}.
It can serve as a stepping stone toward simpler verification approaches.  

A user survey of two voter groups showed that the system was well accepted. Still, some voters considered even \PVV{}'s simple version of verification an additional burden. 
While they could have just seen verification as a waste of their time, people still found the process positive and said that it increased their confidence in the voting system over previous approaches. This first demonstration also confirmed that using the system a few times reduces the difficulty dramatically. 

Adding a small amount of programming would streamline the protocol of \PVV{}, saving time and effort and reducing chances for EA mistakes. Best practices would add EA protocols to avoid malfeasance by the committee chair or the EA (not tested in the demonstrations reported herein).  The EA should have no relationship to the voters or voting organization. Both the EA and the committee chair should be required to show their list of voters, and  compilation of verification and votes, to an impartial university administrator outside the department. This oversight removes them from any suspicion of tampering with the election.  This oversight is newly available in virtual meetings as external personnel are not allowed in the promotion meeting. 

The boardroom voting scenario that \PVV{} speaks to  is quite a different setting from municipal elections but also presents problems of integrity, accuracy, and privacy. Getting people to follow up on anything they do is difficult. Getting even security experts to be willing to take a verification step in a election has been challenging. Modern voting technology insights---software independence, assuring that a person votes, assuring the vote is included---are then all valuable in this special kind of election. The virtual interactions of the pandemic became a forum to learn better how verification can be communicated and acted on. 

One of the main results of this paper is the idea that many kinds of elections occur, with very different requirements, and can be improved with common off-the-shelf technology systems in simple ways. These different requirements show the usefulness of having a range of voting options for various applications, including ones with different trade-offs between simplicity, ease of use, privacy, and outcome integrity. 
Our experiment shows that software independence can be made practical even in a world where computers collect votes. \PVV{} is an easy-to-implement system that can greatly reduce problems over other boardroom elections options.

%%%%%%%%%%%%%%%%%%%%%%%%%%%%%%%%%%%%%%%%%%%%%%%%%%%%%%%%%%%%%%%%%%%%%%%%%%%

\newpage
\section*{Acknowledgments}
\small
{
We thank Rebecca Dongarra and Curtis Menyuk 
for helpful comments.
Alan Sherman was supported in part by
the National Science Foundation under SFS grants DGE-1753681, and 1819521,
and by the U.S. Department of Defense under 
CySP grants H98230-19-1-0308 and H98230-20-1-0384.
Enka Blanchard received support from the French PIA project ``Lorraine Université d'Excellence,'' reference ANR-15-IDEX-04-LUE. 

%%%%%%%%%%%%%%%%%%%%%%%%%%%%%%%%%%%%%%%%%%%%%%%%%%%%%%%%

\begin{comment}
\bigskip \noindent
{\it Preliminary draft ({\today}). To be submitted to SOUPS 2021. There is a 12-page limit (excluding refs and appendix), but shorter papers are encouraged. Registration is due Feb 18 and submissions due Feb 25. Notifications on may 21st (except for early reject).}
\bigskip
\end{comment}

\bibliographystyle{plain}
\bibliography{CATS-bib}

\begin{thebibliography}{10}

\bibitem{adida2008helios}
Ben Adida.
\newblock Helios: Web-based open-audit voting.
\newblock In {\em USENIX Security Symposium}, volume~17, pages 335--348, 2008.

\bibitem{adida2009electing}
Ben Adida, Olivier De~Marneffe, Olivier Pereira, Jean-Jacques Quisquater,
  et~al.
\newblock Electing a university president using open-audit voting: Analysis of
  real-world use of {H}elios.
\newblock {\em EVT/WOTE}, 9(10), 2009.

\bibitem{blanchard2020boardroom}
Enka Blanchard, Ted Selker, and Alan~T. Sherman.
\newblock Boardroom voting: Verifiable voting with ballot privacy using
  low-tech cryptography in a single room.
\newblock \url{https://arxiv.org/abs/2007.14916}, 2020.

\bibitem{Blanchard2018Improvingsecurityusability}
N.~Blanchard, Cl{\'{e}}ment Malaingre, and Ted Selker.
\newblock Improving security and usability of passphrases with guided word
  choice.
\newblock In {\em 34th Annual Computer Security Applications Conference,
  {ACSAC}}, pages 723--732, 2018.

\bibitem{Bonneau2012Linguisticpropertiesmulti}
Joseph Bonneau and Ekaterina Shutova.
\newblock Linguistic properties of multi-word passphrases.
\newblock In {\em International Conference on Financial Cryptography and Data
  Security}, pages 1--12. Springer, 2012.

\bibitem{carback2010scantegrity}
Richard Carback, David Chaum, Jeremy Clark, John Conway, Aleksander Essex,
  Paul~S Herrnson, Travis Mayberry, Stefan Popoveniuc, Ronald~L Rivest, Emily
  Shen, Alan~T Sherman, and Poorvi~L Vora.
\newblock Scantegrity {II} municipal election at {T}akoma {P}ark: The first e2e
  binding governmental election with ballot privacy.
\newblock In {\em Proceedings of {USENIX} Security}. USENIX Association, 2010.

\bibitem{cardillo2018threat}
Anthony Cardillo and Aleksander Essex.
\newblock The threat of {SSL/TLS} stripping to online voting.
\newblock In {\em International Joint Conference on Electronic Voting}, pages
  35--50. Springer, 2018.
\newblock \url{https://whisperlab.org/blog/2016/security-analysis-of-helios }.

\bibitem{chang2016cloudier}
Nicholas Chang-Fong and Aleksander Essex.
\newblock The cloudier side of cryptographic end-to-end verifiable voting: a
  security analysis of {H}elios.
\newblock In {\em Proceedings of the 32nd Annual Conference on Computer
  Security Applications}, pages 324--335, 2016.
\newblock
  \url{https://www.researchgate.net/profile/David_Duenas-Cid/publication/327980266_Third_International_Joint_Conference_on_Electronic_Voting_E-Vote-ID_2018_TUT_Press_Proceedings/links/5bd99588299bf1124fafaba2/Third-International-Joint-Conference-on-}
  \url{Electronic-Voting-E-Vote-ID-2018-TUT-Press- }
  \url{Proceedings.pdf#page=337 }.

\bibitem{votexx}
David Chaum, Richard Carback, Mario Yaksetig, Jeremy Clark, Mahdi Nejadgholi,
  Alan~T. Sherman, Chao Liu, Filip Zagórski, and Bart Preneel.
\newblock Votexx {P}roject.
\newblock \url{https://votexx.org/}, 2020.

\bibitem{Cortier2019Beleniossimpleprivate}
V{\'e}ronique Cortier, Pierrick Gaudry, and St{\'e}phane Glondu.
\newblock Belenios: a simple private and verifiable electronic voting system.
\newblock In {\em Foundations of Security, Protocols, and Equational
  Reasoning}, pages 214--238. Springer, 2019.

\bibitem{hao2016real}
Feng Hao and Peter~YA Ryan.
\newblock {\em Real-World Electronic Voting: Design, Analysis and Deployment}.
\newblock CRC Press, 2016.

\bibitem{helios-attacks-and-defense}
Helios.
\newblock Attacks and defense.
\newblock \url{https://documentation.heliosvoting.org/attacks-and-defenses },
  2011.

\bibitem{javani2020bvot}
Farid Javani and Alan~T Sherman.
\newblock {BVOT}: Self-tallying boardroom voting with oblivious transfer.
\newblock {\em arXiv preprint arXiv:2010.02421}, 2020.
\newblock \url{https://arxiv.org/abs/2010.02421}.

\bibitem{Karayumak2011Userstudyimproved}
F.~{Karayumak}, M.~{Kauer}, M.~M. {Olembo}, T.~{Volk}, and M.~{Volkamer}.
\newblock User study of the improved helios voting system interfaces.
\newblock In {\em 2011 1st Workshop on Socio-Technical Aspects in Security and
  Trust (STAST)}, pages 37--44, Sep. 2011.

\bibitem{Miller1956magicalnumberseven}
George~A. Miller.
\newblock The magical number seven, plus or minus two: Some limits on our
  capacity for processing information.
\newblock {\em Psychological Review}, 63(2):81, 1956.

\bibitem{rivest2008notion}
Ronald~L Rivest.
\newblock On the notion of ‘software independence’in voting systems.
\newblock {\em Philosophical Transactions of the Royal Society A: Mathematical,
  Physical and Engineering Sciences}, 366(1881):3759--3767, 2008.

\bibitem{smyth2018ballot}
Ben Smyth.
\newblock Ballot secrecy: {S}ecurity definition, sufficient conditions, and
  analysis of {H}elios.
\newblock Technical report, Cryptology ePrint Archive, Report 2015/942, 2018.
\newblock \url{https://eprint.iacr.org/2015/942.pdf }.

\bibitem{umbchandbook}
UMBC.
\newblock Faculty {H}andbook.
\newblock \url{https://provost.umbc.edu/policies/faculty-handbook/}.

\bibitem{Vassil2016diffusioninternetvoting.}
Kristjan Vassil, Mihkel Solvak, Priit Vinkel, Alexander~H. Trechsel, and
  Michael~R. Alvarez.
\newblock {The diffusion of Internet voting. Usage patterns of internet voting
  in Estonia between 2005 and 2015}.
\newblock {\em Government Information Quarterly}, 33(3):453--459, 2016.

\bibitem{box}
Box.
\newblock \url{https://www.box.com }.

\bibitem{docusign}
Docusign.
\newblock \url{ https://www.docusign.com/ }.

\bibitem{helios}
Helios: Trust the vote.
\newblock \url{https://heliosvoting.org/}.

\bibitem{rw1}
random word generator.
\newblock \url{https://randomwordgenerator.com/}.

\bibitem{rw2}
word generator.
\newblock \url{https://www.wordgenerator.net/}.

\end{thebibliography}
}
\appendix
\normalsize
\section{Sample Ballot Form}
\includegraphics[width=\columnwidth]{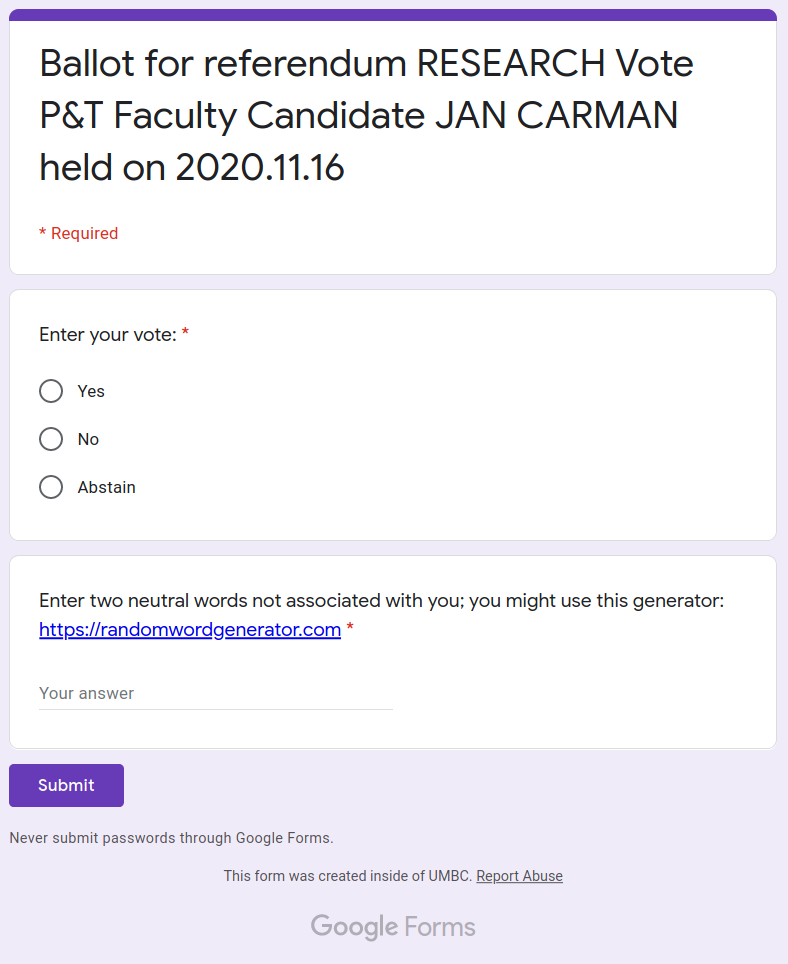}
\label{sec:ballot-ex}

\vspace*{2in} % to set in different cols

\section{Sample Verification Form}
\includegraphics[width=\columnwidth]{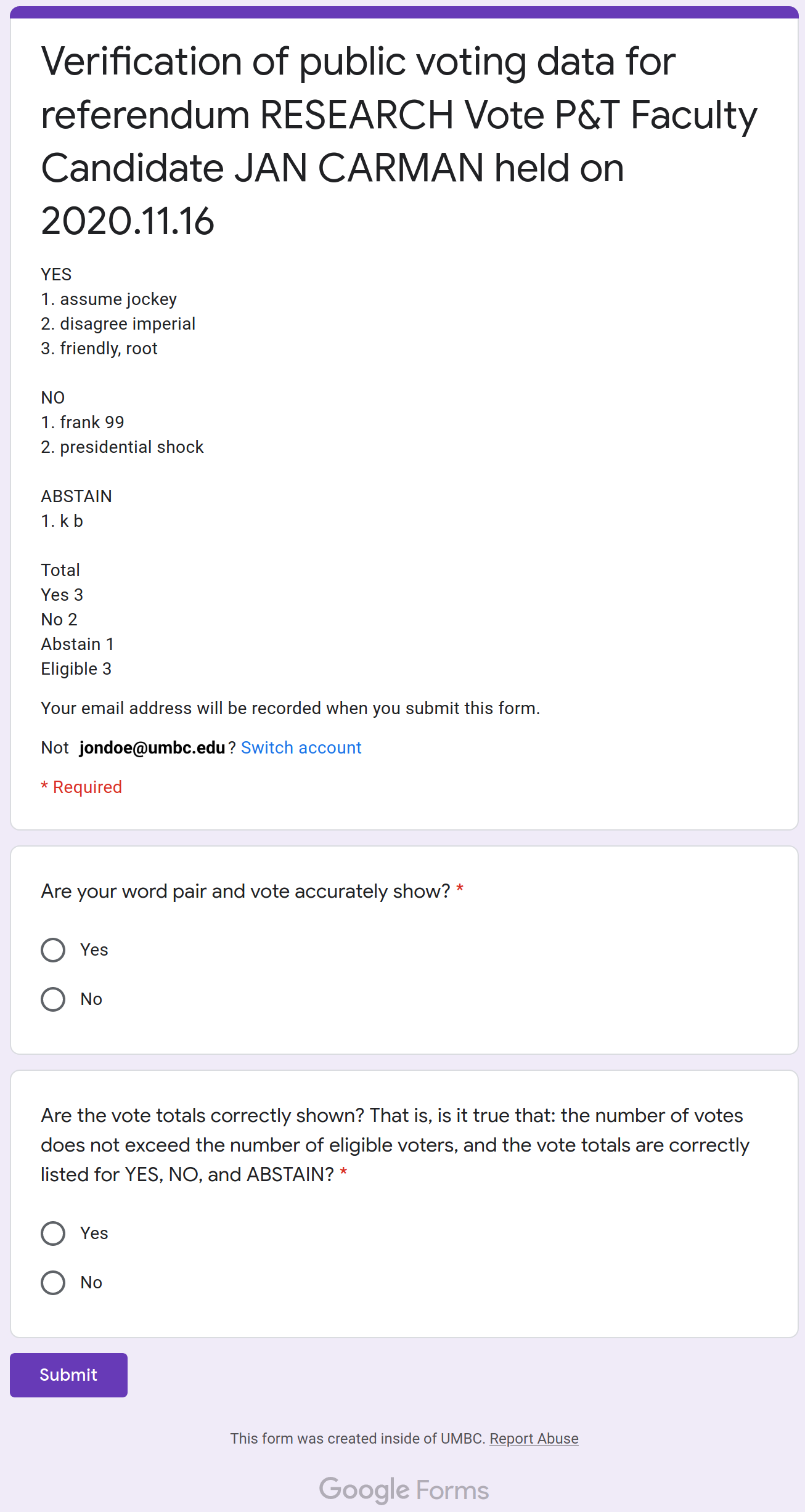}
\label{sec:verification-ex}

%\section{Survey Forms}\label{sec:surveyform}

%%%%%%%%%%%%%%%%%%%%%%%%%%%%%%%%%%%%%%%%%%%%%%%%%%%%%%%%%

%-------------------------------------------------------------------------------

%%%%%%%%%%%%%%%%%%%%%%%%%%%%%%%%%%%%%%%%%%%%%%%%%%%%%%%%%%%%%%%%%%%%%%%%%%%%%%%%
\end{document}